\documentstyle[graphicx,aps,multicol]{revtex}
\draft

\begin{document}

\title{Caloric curves of atomic nuclei and other small systems}
\author{A.~Schiller\footnote{Electronic address: Andreas.Schiller@llnl.gov}}
\address{Lawrence Livermore National Laboratory, L-414, 7000 East Avenue, 
Livermore CA-94551}
\author{M.~Guttormsen, M.~Hjorth-Jensen, J.~Rekstad, and S.~Siem}
\address{Department of Physics, University of Oslo, N-0316 Oslo, Norway}
\maketitle

\begin{abstract}
Caloric curves have traditionally been derived within the microcanonical 
ensemble via $\frac{\partial S}{\partial E}=\frac{1}{T}$ or within the 
canonical ensemble via $E=T^2\frac{\partial \ln Z}{\partial T}$. In the 
thermodynamical limit, i.e., for large systems, both caloric curves give the 
same result. For small systems like nuclei, the two caloric curves are in 
general different from each other and neither one is reasonable. Using 
$\frac{\partial S}{\partial E}=\frac{1}{T}$, spurious structures like negative
temperatures and negative heat capacities can occur and have indeed been 
discussed in the literature. Using $E=T^2\frac{\partial \ln Z}{\partial T}$ a 
very featureless caloric curve is obtained which generally smoothes too much 
over structural changes in the system. A new approach for caloric curves based 
on the two-dimensional probability distribution $P(E,T)$ will be discussed.
\end{abstract}

\pacs{PACS number(s): 05.20.Gg, 05.70.Fh, 21.10.Ma, 24.10.Pa}

\begin{multicols}{2}

\section{Introduction}

A common misconception in phenomenological thermodynamics is the notion that 
the caloric curve is given by 
\begin{equation}
\frac{\partial S}{\partial E}=\frac{1}{T}
\label{eq:mcc}
\end{equation}
in the microcanonical ensemble, and by 
\begin{equation}
E=T^2\frac{\partial\ln Z}{\partial T}
\label{eq:ccc}
\end{equation}
in the canonical ensemble.\footnote{Throughout this work we will set the 
Boltzmann constant $k_{\mathrm{B}}=1$.} This is only true in the 
thermodynamical limit, where both caloric curves coincide, but it is certainly 
not true for small systems like atomic nuclei where the two curves can differ 
dramatically from each other \cite{MB99}.

Closely connected with this error is the believe that the temperature in the 
microcanonical ensemble is \em defined \rm by Eq.\ (\ref{eq:mcc}). At this 
point, we would therefore like to emphasize that the temperature scale is 
defined by the triple point of water and several secondary standards. 
Temperatures themselves are defined by a measurement process, i.e., by a 
comparison of an unknown temperature to temperature standards. Since this 
measurement process requires the exchange of energy between the system under 
study and a thermometer, we can immediately see that the concept of a 
temperature can conflict with the concept of a microcanonical ensemble which 
is, per definition, closed for energy exchange. The solution to this problem 
for large system is the incorporation of a relatively small thermometer into 
the microcanonical system. Only for this case it has been shown experimentally 
that the definition of the temperature coincides with Eq.\ (\ref{eq:mcc}). For 
a small microcanonical system, no such solution exists and the temperature of 
such a system has to remain undefined, since it cannot be measured. Therefore,
for any small system, temperature has always to be introduced by a coupling of 
the system under study to a large heat bath.

\section{Traditional caloric curves}

As starting point for the discussion in this work, we take the probability $P$
of a system to have the energy $E$ for a given temperature $T$ which is imposed
by a large heat bath
\begin{equation}
P(E,T)=\frac{\Omega(E)\exp(-E/T)}{Z(T)}.
\end{equation}
Here, $\Omega(E)$ is the multiplicity of states with energy $E$, and $Z(T)$ is
the canonical partition function
\begin{equation}
Z(T)=\int_0^\infty\Omega(E^\prime)\exp(-E^\prime/T)\,{\mathrm{d}}E^\prime.
\label{eq:laplace}
\end{equation}
In the derivation of this probability distribution, one usually takes advantage
of Eq.\ (\ref{eq:mcc}) applied to the heat bath. Since the heat bath is thought
to be approximately in the thermodynamical limit, the use of Eq.\ 
(\ref{eq:mcc}) is justified and will not conflict with our claim in the present
work that for small systems, Eq.\ (\ref{eq:mcc}) will become unphysical.

Now, for convenience, it is often easier to use the logarithm of the 
probability $P$ than the probability itself. We therefore define
\begin{equation}
A(E,T)=\ln P(E,T).
\end{equation}
In the thermodynamical limit, for a given $T$, the most probable value of $E$ 
which we will denote $\widehat{E}$ throughout this work is determined by the 
condition
\begin{equation}
\left.\frac{\partial A(E,T)}{\partial E}\right|_{E=\widehat{E}}\equiv 
A_E(\widehat{E},T)=0.
\end{equation}
Simple algebra shows that this condition is equivalent to
\begin{equation}
\left.\frac{\partial S(E)}{\partial E}\right|_{E=\widehat{E}}=\frac{1}{T},
\end{equation}
where $S(E)=\ln\Omega(E)$ is the microcanonical entropy. Here, we have 
recovered the traditional expression for the caloric curve in the 
microcanonical ensemble. However, a word of caution is necessary: for small 
systems, the above condition does not necessarily give always the most probable
value $\widehat{E}$, but can, under certain circumstances, yield the locally 
least probable value of $E$, denoted $\check{E}$. We will later give an example
for this claim. It is only in the thermodynamical limit that always the most 
probable value of $E$ is obtained. In this limit, the construction of the 
microcanonical caloric curve is therefore equivalent to finding $\widehat{E}$ 
for a given $T$.

Alternatively, we can search for the most probable value of $T$ (denoted 
$\widehat{T}$) for a given $E$. In the thermodynamical limit, this value is 
obtained through the condition 
\begin{equation}
\left.\frac{\partial A(E,T)}{\partial T}\right|_{T=\widehat{T}}\equiv 
A_T(E,\widehat{T})=0
\end{equation}
which is equivalent to 
\begin{equation}
E=\widehat{T}^2\left.\frac{\partial\ln 
Z(T)}{\partial T}\right|_{T=\widehat{T}}.
\end{equation}
Here, we have recovered the traditional expression for the caloric curve in the
canonical ensemble.

Let us now calculate the average energy 
\begin{equation}
\langle E\rangle=\int_0^\infty\,E^\prime\,P(E^\prime,T)\,{\mathrm{d}}E^\prime
\end{equation}
of the distribution $P$ for a given temperature $T$. Again, simple algebra 
yields
\begin{equation}
\langle E\rangle=T^2\frac{\partial \ln Z(T)}{\partial T}
\end{equation}
which is the same result as before. This means that the construction of the 
canonical caloric curve is equivalent to:
\begin{itemize}
\item finding $\langle E\rangle$ for a given temperature $T$
\item finding $\widehat{T}$ for a given energy $E$.
\end{itemize}
The first feature makes the canonical curve more robust and therefore more 
attractive than the microcanonical curve.

In general, caloric curves are constructed by replacing the distribution of 
energies (temperatures) for a given temperature (energy) by a special value of
this distribution, often the most probable value or the mean value. This gives 
satisfactory results in the thermodynamical limit, where the distributions are 
very sharp, almost $\delta$-like, and where therefore the mean value and the 
most probable value will coincide. This causes also that the traditional 
microcanonical and canonical caloric curves coincide in the thermodynamical 
limit. For small systems, however, the probability distribution $P$ can be 
quite flat and exhibit more than one maximum. Thus, a one-by-one relation of 
temperature and energy as represented by a caloric curve can only serve as an 
approximate concept. Further, as shown above, the approximations in deriving a 
caloric curve for small systems are manifested differently in the 
microcanonical and canonical ensemble, i.e.\ by finding $\widehat{E}$ (or 
sometimes $\check{E}$) for a given $T$ in the microcanonical ensemble or by 
finding $\widehat{T}$ for a given $E$ in the canonical ensemble (or 
alternatively, by finding $\langle E\rangle$ for a given $T$). Thus, the two 
resulting caloric curves can, in general, be vastly different for the same 
(small) system. 

The physical reason behind the inherently approximative nature of the concept 
of caloric curves for a small system is the Laplace transformation which 
connects the two statistical ensembles as shown in Eq.\ (\ref{eq:laplace}). 
From a mathematical point of view, this transformation is not unlike the 
Fourier transformation connecting the coordinate and momentum space in quantum 
mechanics. In analogy to quantum mechanics, we can therefore regard the 
microcanonical ensemble as the 'energy representation' of the system, since the
control parameter is the energy, while the temperature is fluctuating. The 
canonical ensemble can then be regarded as the 'temperature representation', 
since here, the temperature is the control parameter while the energy is 
fluctuating. Further, just as the exact knowledge of momentum (or coordinate) 
implies, according to Heisenberg, large uncertainties in the other quantity, in
thermodynamics of small systems, a fixed value of $T$ (imposed by a heat bath) 
implies large fluctuations in $E$, and a fixed $E$ (as in the microcanonical 
ensemble) implies an ill-defined $T$ due to fluctuations. Only in the 
thermodynamical limit, the relative fluctuations will become negligibly small 
and the caloric curve will again become an exact concept, just as in classical 
mechanics, an object can again have a definite coordinate and momentum.

Acknowledging the approximative nature of the concept of caloric curves, we 
will, in the following, show that in general neither the microcanonical nor the
canonical ensemble actually gives a good approximation for the caloric curve. 
Therefore, a new method will be proposed to construct more meaningful caloric 
curves for small systems.

\section{Shortcomings of traditional caloric curves in small systems}

\begin{figure}\centering
\includegraphics[totalheight=17.9cm]{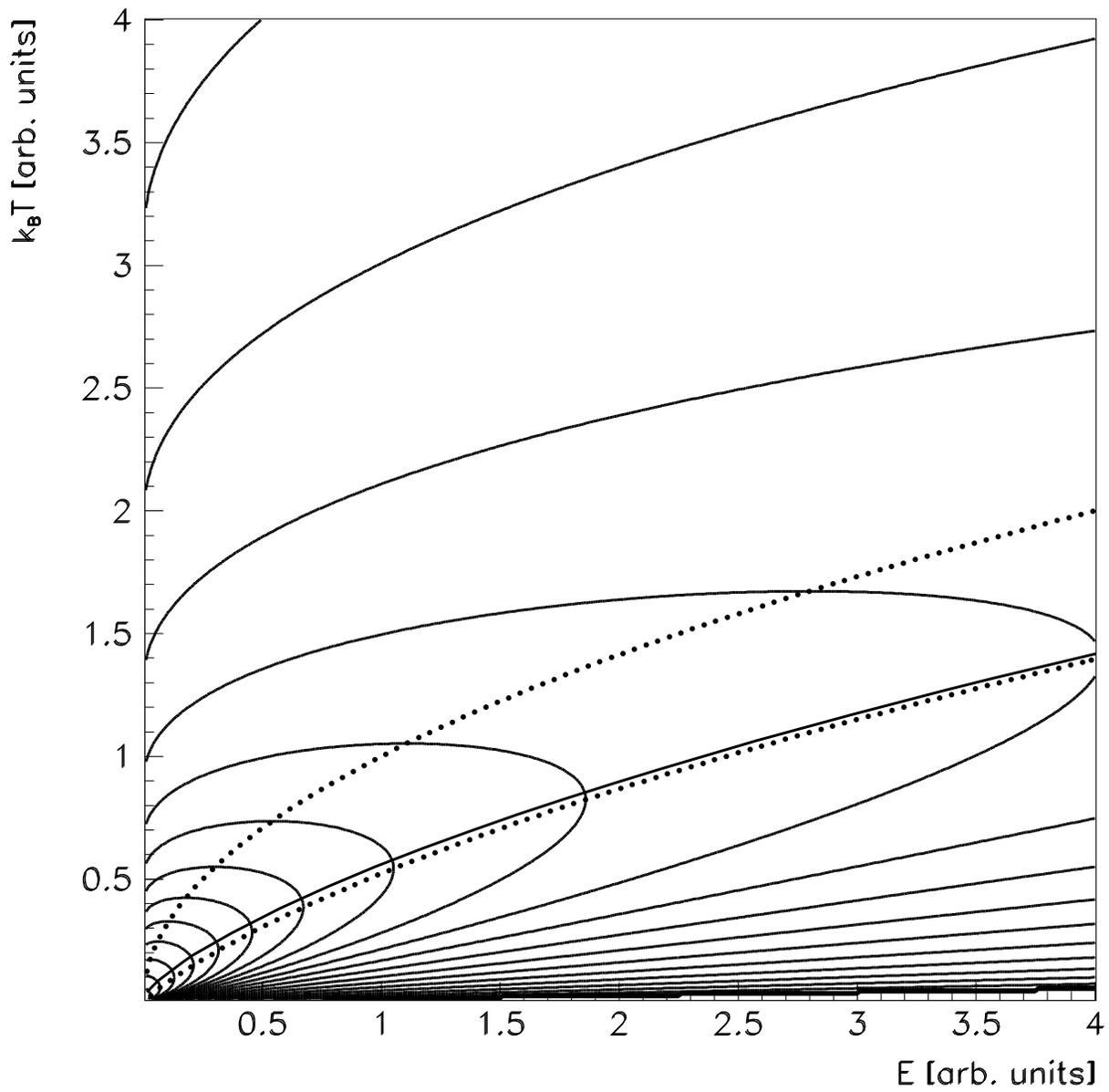}
\caption{Same as previous two figures but for a simple Fermi-gas level 
density.}
\label{fig:netbethe}
\end{figure}

\begin{figure}\centering
\includegraphics[totalheight=8.9cm]{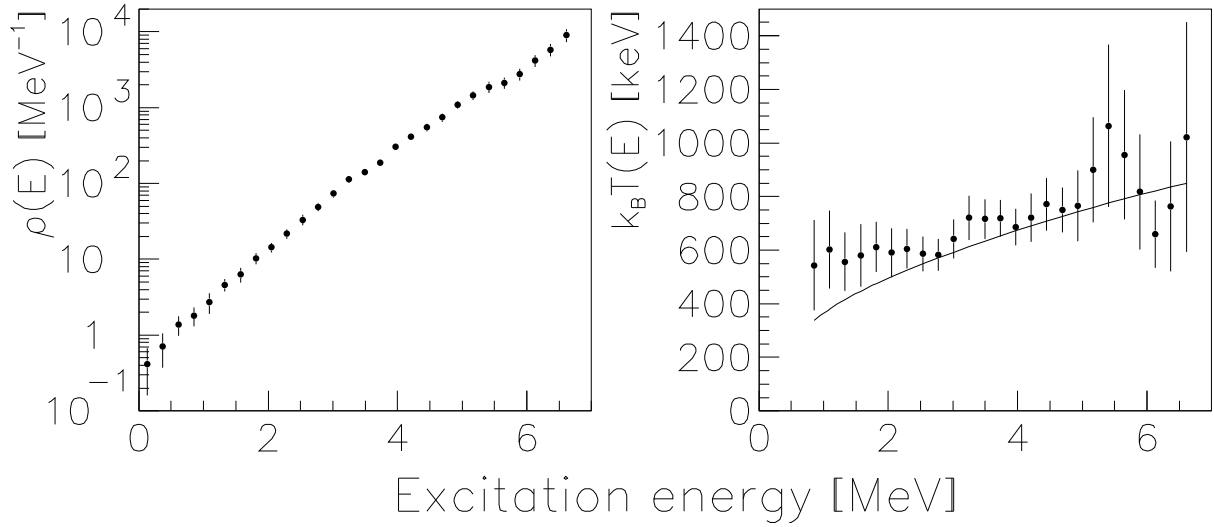}
\caption{Left-hand panel: experimental level density for the nucleus $^{96}$Mo 
obtained by the method described in \protect\cite{ST03}. Right-hand panel: 
microcanonical (data points with error bars) and canonical (solid line) caloric
curves for this nucleus. For the purpose of the canonical caloric curve, the 
experimental level density had to be extrapolated by a Fermi-gas expression in 
order to evaluate Eq.\ (\ref{eq:laplace}). The qualitative agreement between 
this figure and Fig.\ \ref{fig:netbethe} is excellent.}
\label{fig:mo96}
\end{figure}

First, we will investigate a simple example where the multiplicity of states is
proportional to $E^n$. Simple algebra shows that the microcanonical caloric 
curve is given by
\begin{equation}
\widehat{E}=n\,T
\end{equation}
whereas the canonical caloric curve is given by 
\begin{equation}
\widehat{T}=E/(n+1).
\end{equation}
Figure \ref{fig:net1} shows $P(E,T)$ together with the two traditional caloric
curves for the case $n=1$. Obviously, the two curves do not coincide and 
neither seems to represent a good characterization to the probability 
distribution $P$. In this example, the thermodynamical limit is approached by 
increasing $n$. Figure \ref{fig:net5} gives the case for $n=5$ where both 
approaches yield more similar curves and the collapse of the probability 
distribution $P$ into a caloric curve becomes a better approximation, mainly 
because $P$ exhibits much sharper maxima as function of $E$ or $T$. As a third 
example, we take a simple Fermi-gas level-density expression
\begin{equation}
\varrho\propto \exp (2\sqrt{aE})
\end{equation}
which gives the usual
\begin{equation}
\widehat{E}=aT^2
\label{eq:micnuc}
\end{equation}
in the microcanonical ensemble, and the less used, but more accurate
\begin{equation}
E=a\widehat{T}^2+\widehat{T}\left[1+\frac{1}
{2+1\left/x(\widehat{T})\right.}\right]
\label{eq:cannuc}
\end{equation}
with
\begin{equation}
x(T)=\sqrt{aT}e^{aT}\int_{-\infty}^{\sqrt{aT}}e^{-\tau^2}{\mathrm{d}}\tau
\end{equation}
in the canonical ensemble\footnote{For large $\widehat{T}$, this expression 
simplifies to 
$E=a\widehat{T}^2+\frac{3}{2}\widehat{T}+{\mathcal{O}}(\widehat{T}^{-1})$, for 
small $\widehat{T}$ (i.e., large $\widehat{\beta}=1/\widehat{T}$), one obtains 
$E=\widehat{\beta}^{-1}+\frac{\sqrt{\pi a}}{2}\widehat{\beta}^{-3/2}+
a\,(2-\pi/2)\widehat{\beta}^{-2}+{\mathcal{O}}(\widehat{\beta}^{-5/2})$.} 
(see Fig.\ \ref{fig:netbethe}). An actual example for caloric curves from a 
system which exhibits approximately a Fermi-gas level density is given in Fig.\
\ref{fig:mo96}. 

\begin{figure}\centering
\includegraphics[totalheight=17.9cm]{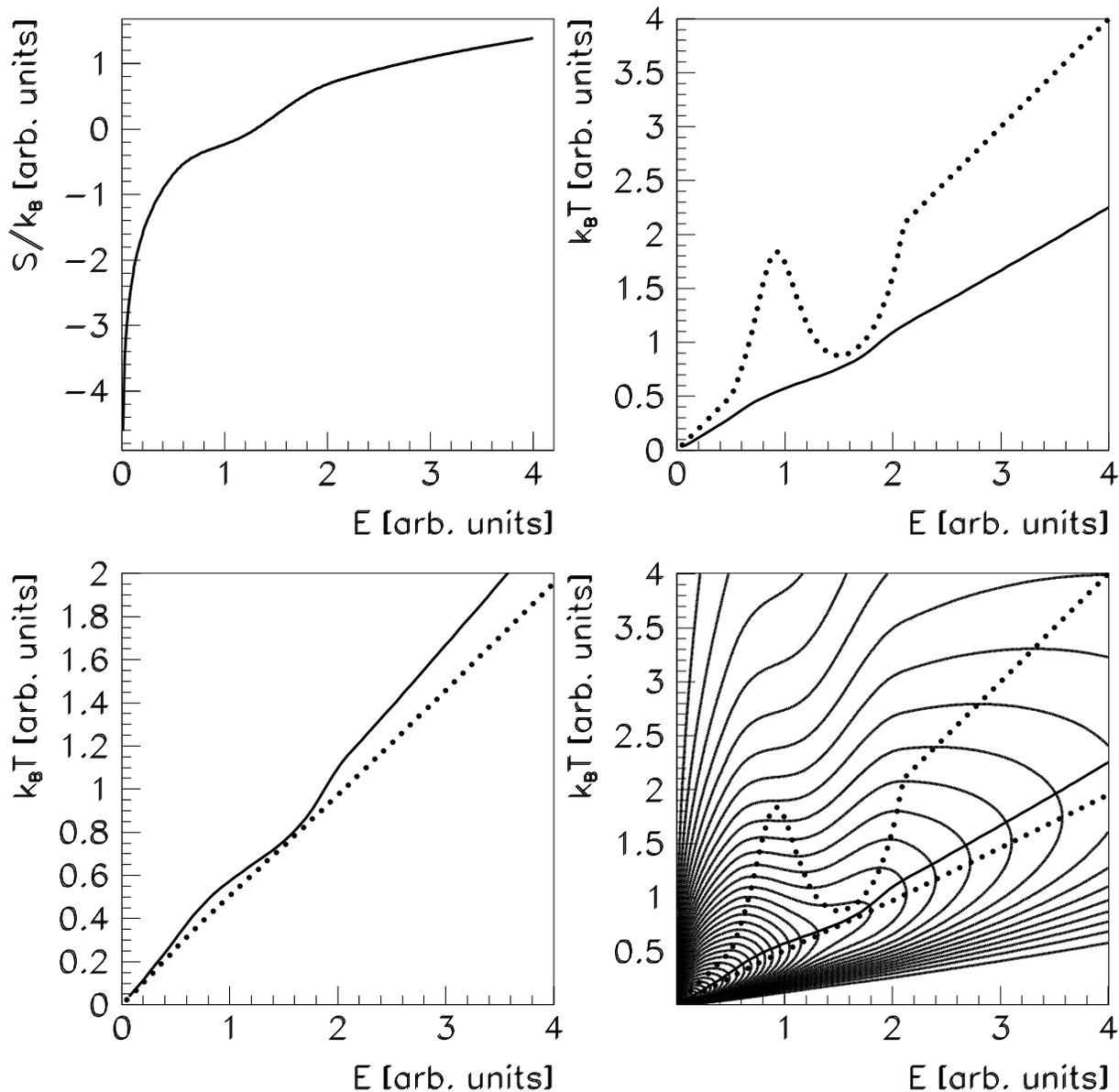}
\caption{Adding a convex intruder to the microcanonical entropy curve (upper 
left-hand panel). The traditional microcanonical caloric curve is the dotted 
line in the upper right-hand panel and displays the often discussed phenomenon 
of a 'backbending', i.e.\ it is multivalued for a region of temperatures. The 
canonical caloric curve is the dotted line in the the lower left-hand panel. It
is very smooth and fails to reflect the structural change of the system 
indicated by the dip in the entropy curve. The contour plot of the probability 
distribution on the lower right-hand panel shows that in the region of the 
'backbending' the microcanonical caloric curve characterizes $P$ by the locally
least probable value $\check{E}$ for a given $T$. Since it does not make sense 
to replace any distribution by its (locally) least probable value, the 
microcanonical approach to a caloric curve is obviously inappropriate and the 
associated spurious structures like the negative heat capacity are unphysical. 
The dotted lines in this panel are again the traditional caloric curves, while 
the solid line in this panel (and in the two previous panels) are the new 
caloric curve from Eq.\ (\ref{eq:perp}) with the correct asymptotic behavior.}
\label{fig:small}
\end{figure}

\begin{figure}\centering
\includegraphics[totalheight=17.9cm]{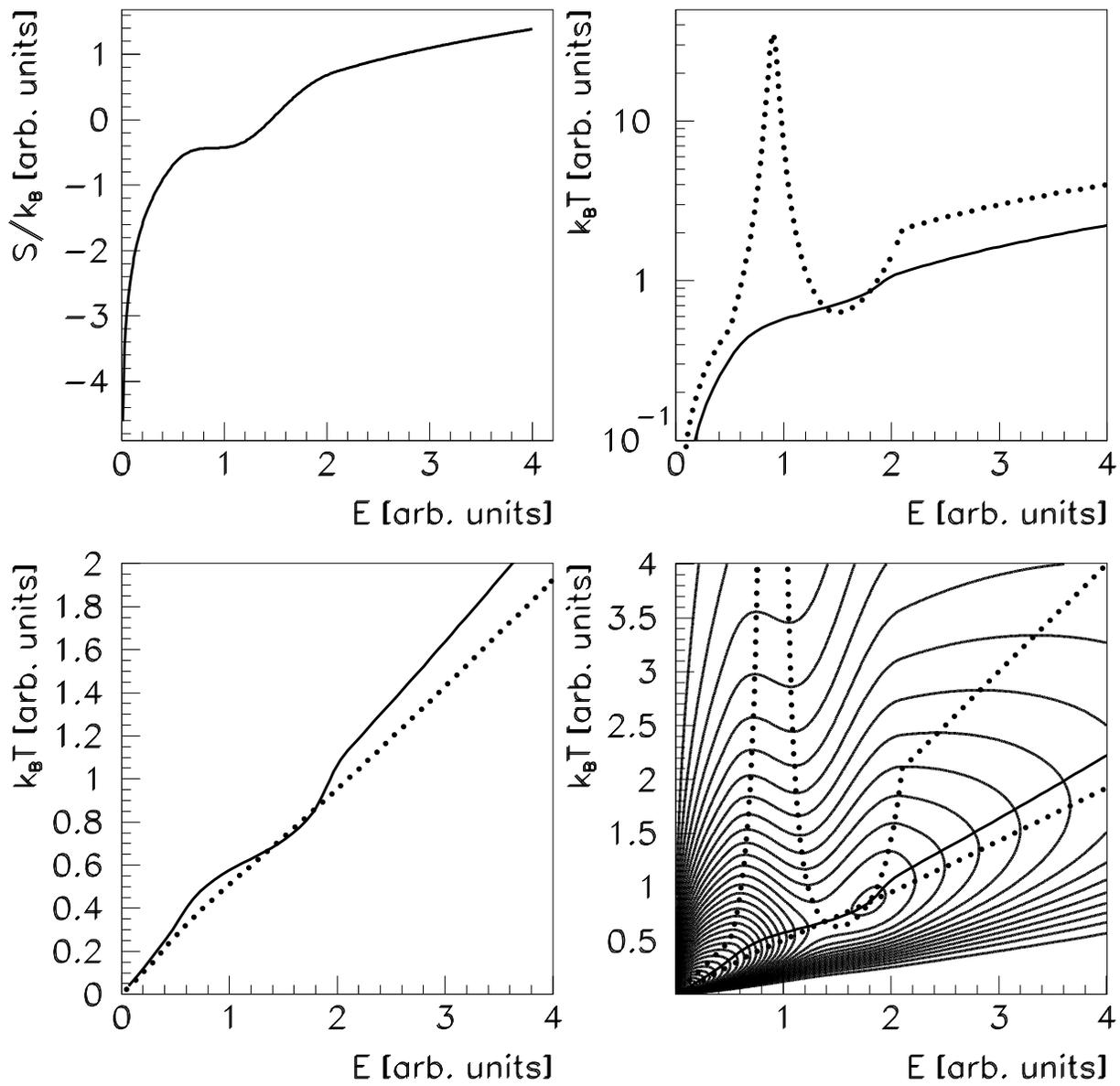}
\caption{Same as previous figure but with a larger dip in the microcanonical 
entropy curve, causing the two caloric curves to cross.}
\label{fig:medium}
\end{figure}

\begin{figure}\centering
\includegraphics[totalheight=8.9cm]{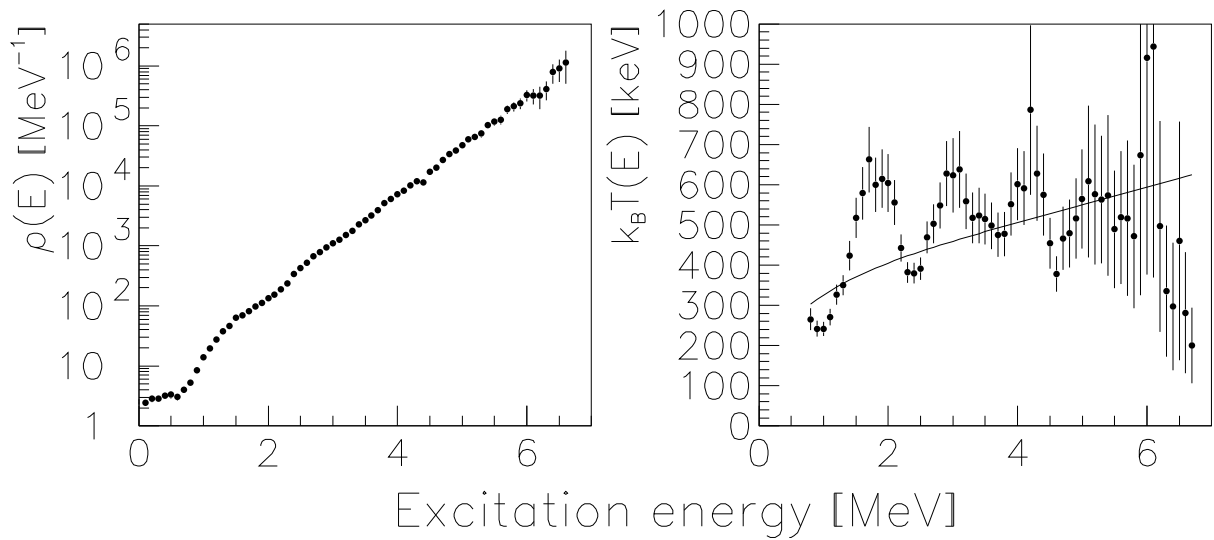}
\caption{Left-hand panel: experimental level density for the nucleus $^{172}$Yb
using the ($^3$He,$\alpha\gamma$) reaction as described in 
\protect\cite{MB99,SB01,VG01}. Right-hand panel: microcanonical (data points 
with error bars) and canonical (solid line) caloric curves for this nucleus. 
The extrapolation of the level density in order to evaluate Eq.\ 
(\ref{eq:laplace}) has been discussed in \protect\cite{SB01}. The experiment 
shows beautifully the possibility of crossing caloric curves for instances 
where the logarithmic level density exhibits convex regions.}
\label{fig:yb172}
\end{figure}

An interesting example is obtained when a convex intruder is present in the 
microcanonical entropy curve. Such a structure is only possible for finite 
systems since for infinite systems, the second law of thermodynamics requires a
concave entropy as function of energy. In the literature, such structures have
been discussed, e.g., for the melting of atomic clusters \cite{SK01}, 
fragmentation of nuclei \cite{Gr97}, and large composite systems like star 
clusters \cite{Th70}. Figure \ref{fig:small} gives an example with a relatively
weak dip in the entropy curve. For this example, the microcanonical caloric 
curve lies always above the canonical caloric curve. In the next example (see 
Fig.\ \ref{fig:medium}), the dip in the entropy curve is made larger, such that
the microcanonical and canonical caloric curves cross each other at two points.
At the two crossing points the probability distribution $P$ has a saddlepoint 
and a local maximum, respectively. However, in general, also here the 
microcanonical caloric curve lies mostly above the canonical caloric curve, in 
agreement with experimental observations \cite{MB99}. Figure \ref{fig:yb172} 
gives an example of an atomic nucleus where the logarithmic level density has 
several convex regions. Those convex intruders indeed cause the microcanonical 
and canonical caloric curves to cross several times. The structural changes in 
the $^{172}$Yb nucleus which cause the convex regions in the microcanonical 
entropy curve in the first place are thought to be the breakup of nucleon 
Cooper pairs \cite{MB99,SB01}.

\begin{figure}\centering
\includegraphics[totalheight=17.9cm]{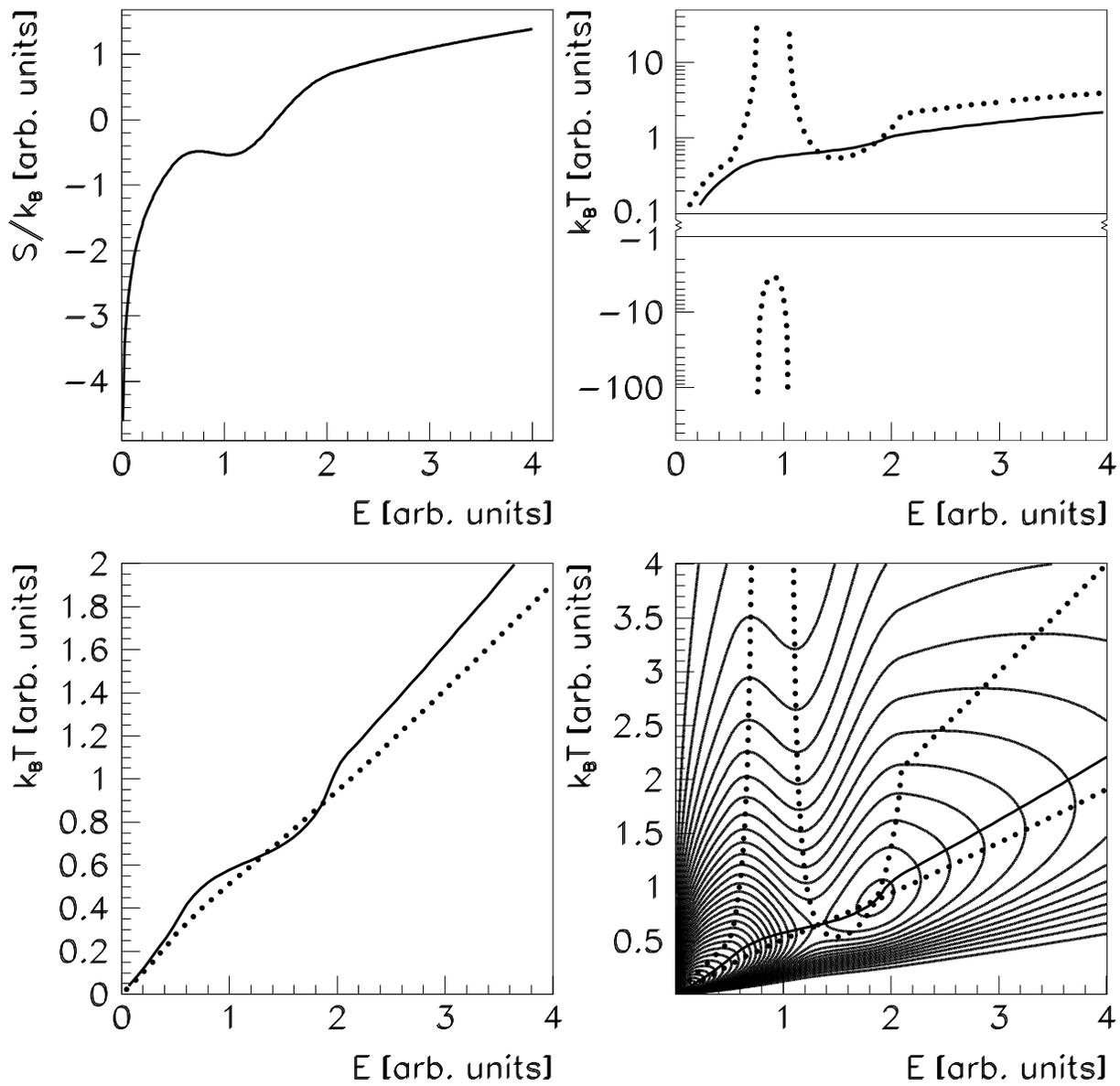}
\caption{Same as Fig.\ \ref{fig:medium} but with an even larger dip in the 
microcanonical entropy curve causing the microcanonical caloric curve to 
exhibit a negative branch for some energies.}
\label{fig:large}
\end{figure}

\begin{figure}\centering
\includegraphics[totalheight=8.9cm]{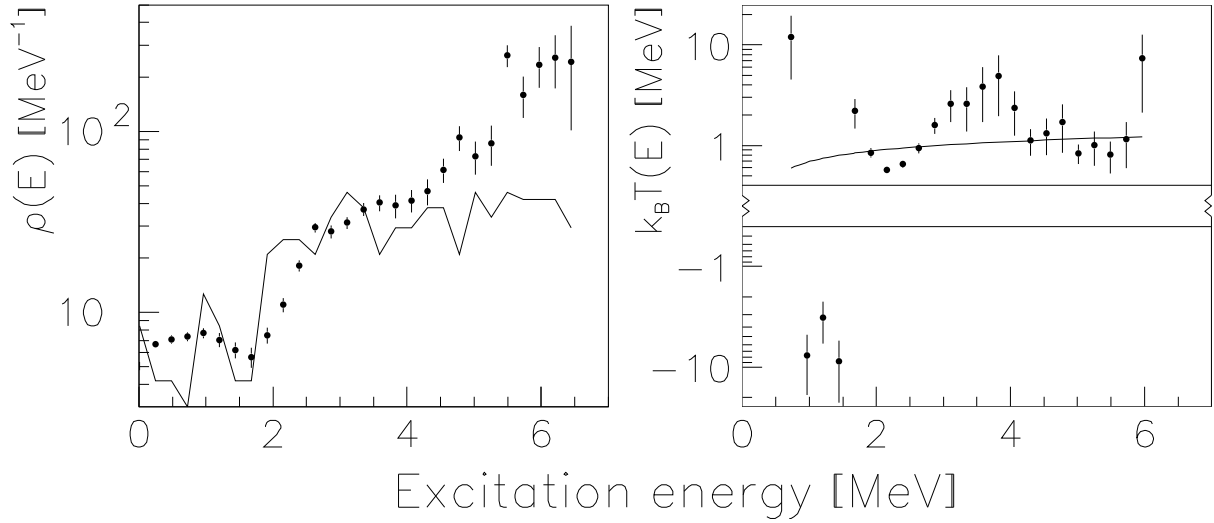}
\caption{Left-hand panel: experimental level density for the nucleus $^{57}$Fe
using the ($^3$He,$^3$He$^\prime\gamma$) reaction as described in 
\protect\cite{ST03}. The jagged line is the level density taken from counting
of discrete levels \protect\cite{FS96}. The binning procedure and fluctuations
in the level spacings at low excitation energies induce larger fluctuations in
this curve compared to the data points which were obtained by statistical 
spectroscopy. The qualitative features of both curves, however, show remarkable
agreement. Right-hand panel: microcanonical (data points with error bars) and 
canonical (solid line) caloric curves for this nucleus \protect\cite{Ta02}. The
local decrease in the logarithmic level density below 2~MeV excitation energy 
creates a negative branch in the microcanonical caloric curve.}
\label{fig:fe57}
\end{figure}

Finally, the dip can be made so large that the entropy locally decreases with 
increasing energy (see Fig.\ \ref{fig:large}). This means that the probability 
distribution $P$ for all large enough $T$ exhibits at least two local maxima 
(and in many cases a local minimum in between), a phenomenon not uncommon for 
small (especially discrete) systems. The negative branch of the microcanonical 
caloric curve is again an unphysical structure induced by the attempt to 
characterize the probability distribution $P$ for a fixed $T$ by an extremum of
$E$. Also this extreme case has been observed in the experimental level density
of the $^{57}$Fe nucleus \cite{Ta02} (see Fig.\ \ref{fig:fe57}). The physical 
origin of the local decrease in the level density of $^{57}$Fe is probably 
simply a fluctuation in the level spacing. The sharp increase in level density 
above 2~MeV excitation energy has again been attributed to the breaking of 
proton Cooper pairs \cite{ST03}.

Until now, the emphasis has been on the unphysical results one obtains when 
using the microcanonical caloric curve. However, also the canonical caloric 
curve does not capture the thermodynamics of the systems under study correctly.
In general, the canonical caloric curve smoothes too much over structural 
changes and remains always featureless, which makes it rather unattractive (see
Figs.\ \ref{fig:small}--\ref{fig:fe57}). To overcome this problem, its 
derivative, the canonical heat capacity 
\begin{equation}
C_V=\left(\frac{\partial \langle E\rangle}{\partial T}\right)_V
\end{equation}
has been used with success to characterize structural changes in real nuclei, 
like the quenching of pairing correlations as function of temperature 
\cite{SB01,LA01,ER00}. The lack of features in the canonical caloric curve 
becomes even clearer for the more extreme examples described in Fig.\ 
\ref{fig:twolev}. There, a two-level system is discussed where the degeneracy 
of the upper level is thrice the degeneracy of the lower level. One could, 
e.g., think of the two levels as a spin 0 ground state and a spin 1 excited 
state.\footnote{For the present model of a two-level system, there is no real
corresponding example from nuclear physics. However, one can easily imagine 
that a similar system could actually be realized in, e.g., atomic physics.} For
the sake of the discussion, a width has been given to both levels such that the
two levels do (upper left-hand panel) or do not (upper right-hand panel) 
overlap. The probability distributions are centered around the level energies 
$E_1$ and $E_2$. While the microcanonical curve is again multivalued and 
exhibits negative branches (not shown in the figure), the canonical caloric 
curve is smooth and reaches the value of $E=(E_1+3\,E_2)/4$ for infinite 
temperatures. Evident from the figure is the fact that at temperatures around 
$T=1.4$ for the lower left-hand panel and $2.4$ for the lower right-hand panel,
the probability for the particle to be located in the lower or upper level 
becomes approximately equal. The canonical caloric curve now still connects 
locations with vertical tangents of the contour lines, i.e., it describes for a
given $E$ the system by the most probable value $\widehat{T}$. On the other 
hand, as can be seen in the lower panels of Fig.\ \ref{fig:twolev}, for 
energies which are rather improbable (left-hand side) or simply cannot be 
realized by the system like $E=1.8$--2.2 (right-hand side), this feature does 
not make sense. Neither does the other feature of the canonical caloric curve 
make sense, i.e., to describe the system by its average energy 
$\langle E\rangle$ for a given $T$, as long as $\langle E\rangle$ coincides 
with values close to the locally least probable value $\check{E}$ (lower 
left-hand panel) or to a completely impossible value of $E$ (lower right-hand 
panel). The canonical caloric curve simply dwells for a too long temperature 
region in the vicinity of improbable or impossible values of $E$. 

\section{New caloric curve}

It is now an interesting question how to define a caloric curve for a small 
system which can overcome all of these problems. In general, the system has of 
course always to be described by the probability distribution, since $P$ is 
most often flat and can exhibit more than one maximum. Therefore, the 
distribution cannot be collapsed to one special value without loosing some of 
the dynamics. On the other hand, caloric curves have been very useful for large
systems and it might prove fruitful to extend the concept of a caloric curve to
small systems as well. The only obvious guidelines for constructing a caloric 
curve is now that it should coincide in the thermodynamical limit to the 
traditional constructions of a caloric curve and that it does, at no point (or
at as few points as possible), collapse the probability distribution to a 
locally least probable value (or to a value close to a locally least probable 
value). 

\begin{figure}\centering
\includegraphics[totalheight=17.9cm]{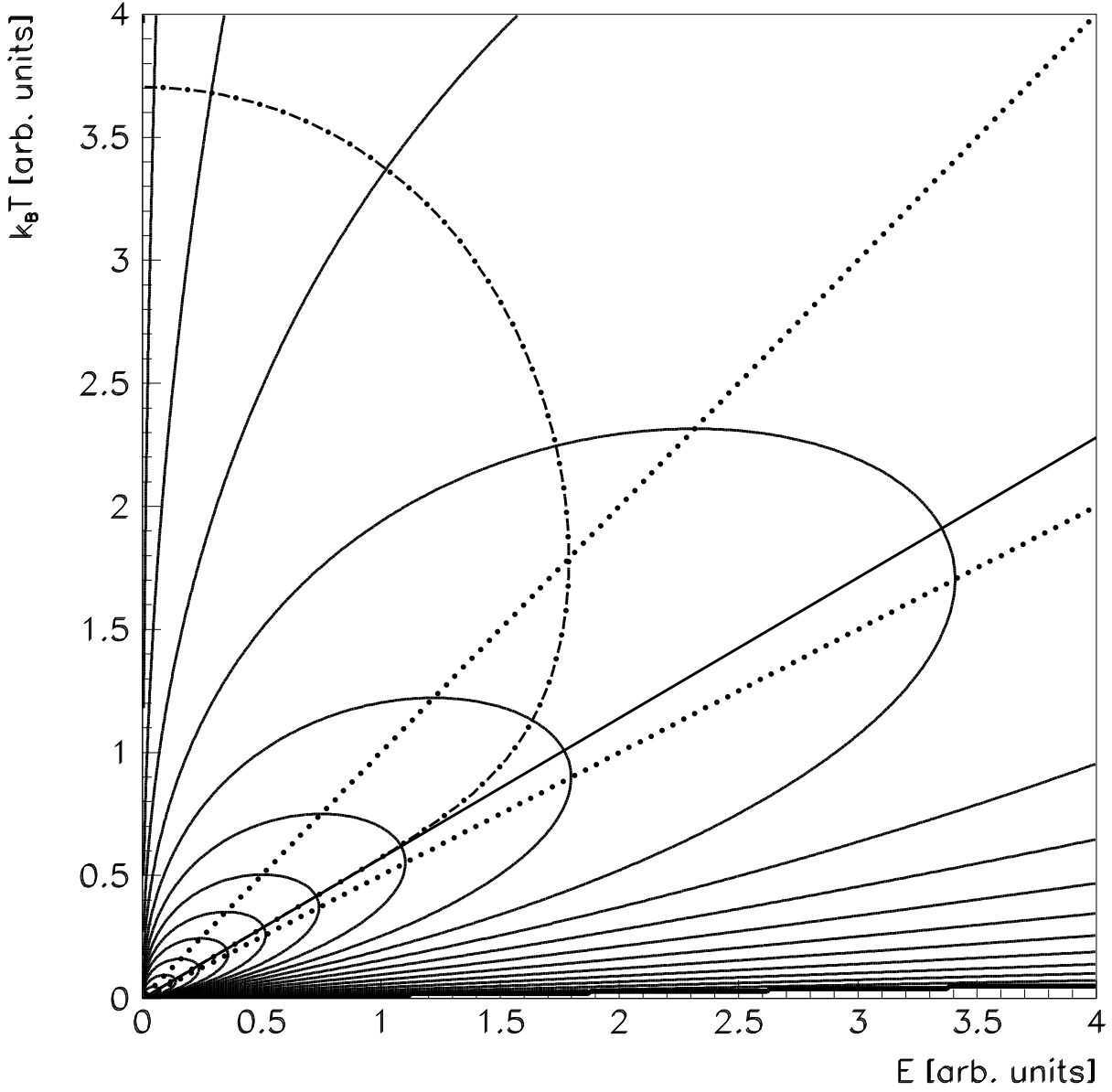}
\caption{Contour plot of the probability distribution $P(E,T)$ for the system 
$\Omega(E)\propto E^n$ with $n=1$. The traditional caloric curves (dotted 
lines) connect points with 
$\left.\frac{\partial P(E,T)}{\partial E}\right|_{E=\widehat{E}}=0$ for the 
microcanonical ensemble and 
$\left.\frac{\partial P(E,T)}{\partial T}\right|_{T=\widehat{T}}=0$ for the 
canonical ensemble. The new approach to caloric curves in a small systems is 
given by the solid line. The dash-dotted line is also a solution to Eq.\ 
(\ref{eq:perp}) but it does not have the correct asymptotic behavior.}
\label{fig:net1}
\end{figure}

\begin{figure}\centering
\includegraphics[totalheight=17.9cm]{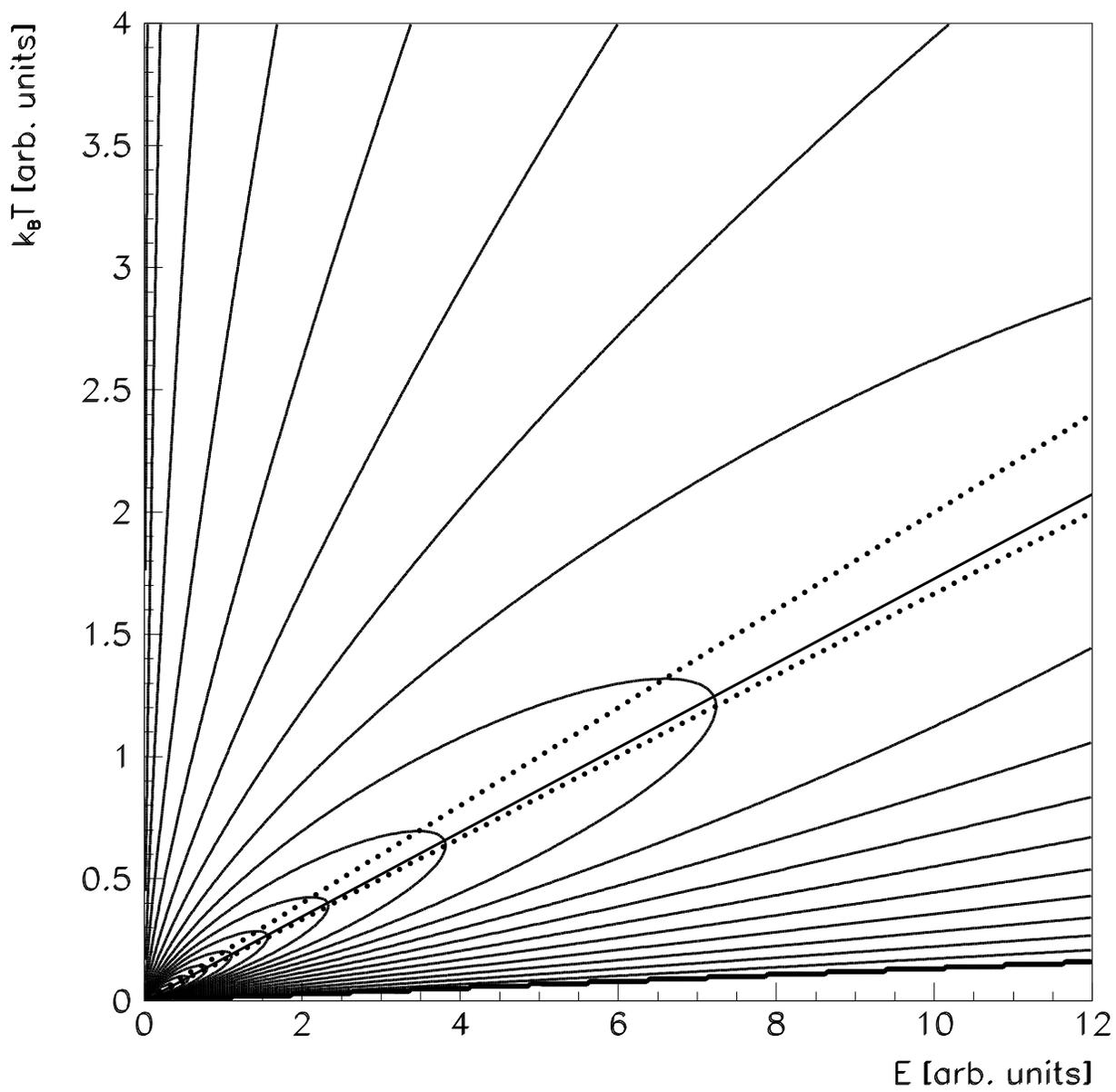}
\caption{Same as previous figure but for $n=5$.}
\label{fig:net5}
\end{figure}

Obviously, one could imagine many different possibilities to define a caloric
curve for small systems which will fulfill the abovementioned conditions. In 
this sense, the approach chosen in this work is certainly not unique, and it 
can be discussed whether it is the best choice for all cases. Our intuitive 
assumptions for the construction of a new caloric curve come from a close 
inspections of Figs.\ \ref{fig:net1} and \ref{fig:net5}. There, we see that for
increasing values of $n$ (which corresponds to approaching the thermodynamical 
limit), the curves of constant probability become more and more pointed and the
points with horizontal and vertical tangents move closer together. Our feeling
is therefore, that an appropriate 'mean' caloric curve is the curve which 
always lies in between the two traditional caloric curves and crosses the 
contour curves of equal probability perpendicularly. Now, the contour curves of
constant probability $P$ are characterized by the differential equation
\begin{equation}
{\mathrm{d}}A=A_E\,{\mathrm{d}}E+A_T\,{\mathrm{d}}T=0.
\end{equation}
The curves perpendicular to those contour curves have a first derivative which
is the negative inverse of the first derivative of the contour curves and are 
thus given by
\begin{equation}
A_T\,{\mathrm{d}}E=A_E\,{\mathrm{d}}T,
\end{equation}
which is equivalent to the differential equation
\begin{equation}
\left(\frac{E}{T^2}-\frac{\partial}{\partial T}\ln Z\right){\mathrm{d}}E=
\left(\frac{\partial}{\partial E}\ln\Omega-\frac{1}{T}\right){\mathrm{d}}T.
\label{eq:perp}
\end{equation}
This equation describes all (infinitely many) curves which perpendicularly 
cross the contour curves of equal probability (one of them is given as the 
dash-dotted line in Fig.\ \ref{fig:net1}), and we therefore have to introduce 
the second part of our assumption in order to single out one solution: the new 
caloric curve should always lie between the two traditional caloric curves. 
This is mathematically done by looking at the asymptotic behavior of the curves
described by Eq.\ (\ref{eq:perp}). The solid lines in Figs.\ 
\ref{fig:net1}--\ref{fig:netbethe} are such curves. They are in general closer 
to the canonical caloric curves, which supports our claim that Eq.\ 
(\ref{eq:cannuc}) is a better representation of the nuclear temperature than 
Eq.\ (\ref{eq:micnuc}), and do not show any surprising features.

The solid line in Fig.\ \ref{fig:small} in the lower right-hand panel is also a
solution of Eq.\ (\ref{eq:perp}) with the correct asymptotic behavior. Again, 
the new caloric curve is rather close to the canonical caloric curve, but it 
shows clearly more features (reflecting the structural change of the system 
under study, which is modeled by the convex entropy intruder). Of course, these
features could again be amplified by taking the derivative of the caloric 
curve, i.e., the heat capacity, as has been done for the canonical caloric 
curve. In Figs.\ \ref{fig:medium} and \ref{fig:large}, the two traditional 
caloric curves actually cross, and the condition that the new caloric curve 
should always lie in between the two traditional caloric curves imposes that 
the new caloric curve also goes through the two crossing points. From a 
mathematical point of view, it is now easier to calculate the new caloric curve
starting from the saddlepoint of the probability distribution, going in both 
directions. The new caloric curve beyond the local maximum in the probability 
distribution, however, is still determined by the correct asymptotic behavior. 
Obviously, the features in the new caloric curve become more visible the larger
the convex intruder in the entropy becomes. Also, the features are always more 
pronounced than in the canonical caloric curve. On the other hand, the new 
caloric curve avoids spurious structures like negative heat capacities (i.e.\ 
'backbending') and branches of negative temperatures, which have been puzzling 
and often led to claims of first-order phase transitions in small systems. 

Unfortunately, the quality of our experimental data on nuclear level densities
does not allow us to compute the new caloric curve according to Eq.\ 
(\ref{eq:perp}). The main problem is the large granularity of the data, i.e., 
the binning into $\sim$120~keV broad excitation energy intervals. Also, the 
error bars of the logarithmic nuclear level density are not sufficiently small 
in order to determine its first derivative with good enough accuracy, as can 
already be seen for the experimental microcanonical caloric curves.

\begin{figure}\centering
\includegraphics[totalheight=17.9cm]{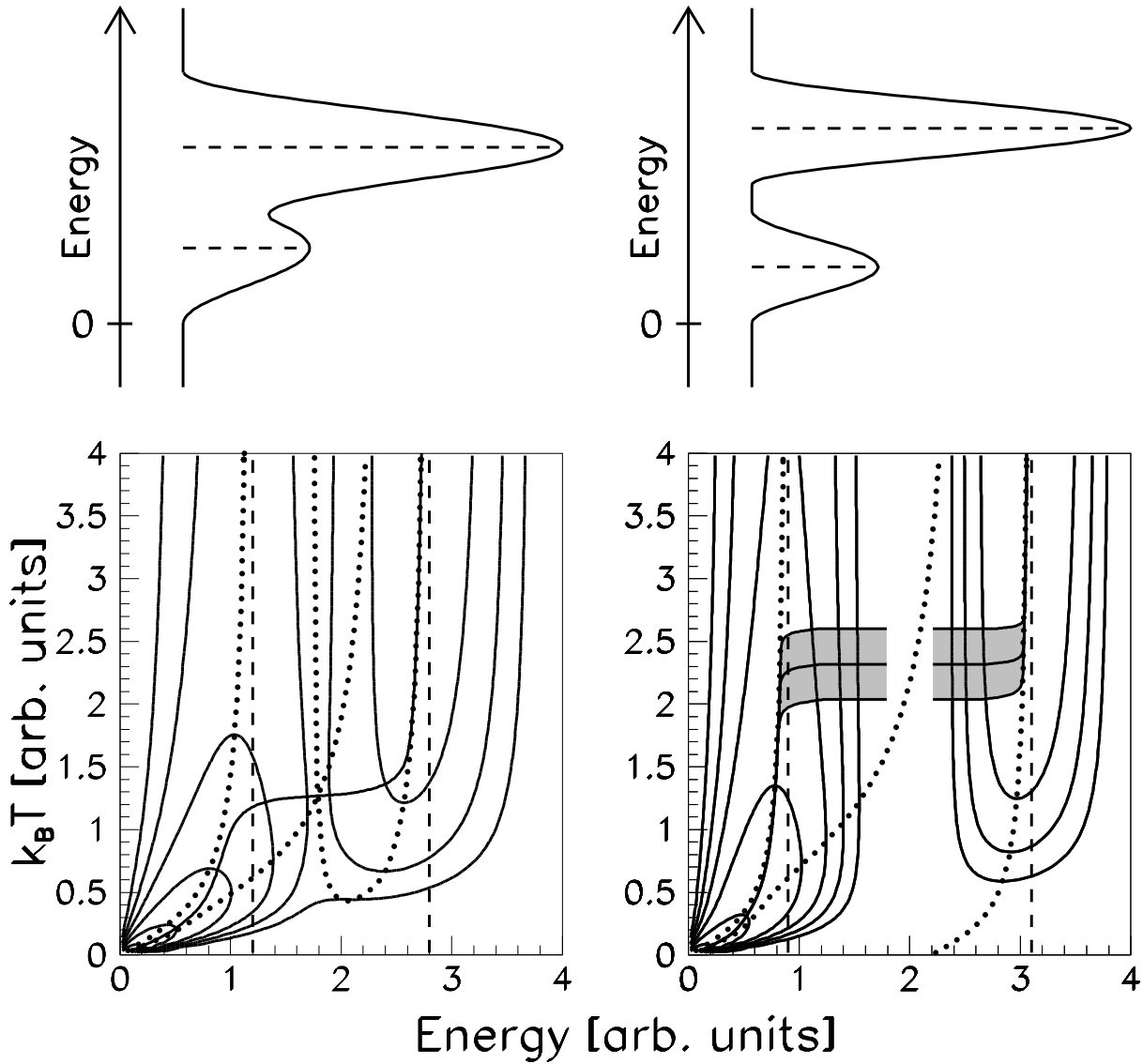}
\caption{Upper panels: two level systems with finite widths. Lower panels: 
contour plots of the probability distributions for the respective systems 
(solid curves), including the two traditional caloric curves (dotted lines), 
the level energies (dashed lines), and solutions to Eq.\ (\ref{eq:perp}) (solid
lines). For some values of $T$, the canonical caloric curve can take values of 
$E$ which are completely impossible (right-hand side) since they do not 
correspond to any allowed energy in the level scheme.}
\label{fig:twolev}
\end{figure}

Finally, the lower left-hand panel of Fig.\ \ref{fig:twolev} shows that the new
caloric curve tends to avoid the region of improbable values of $E$ for a given
$T$ by sweeping fast from $E_1$ to the $E_2$ within a very small temperature 
interval. This is also an example where the new caloric curve for most values 
of $T$ actually comes closer to the microcanonical caloric curve than to the 
canonical caloric curve. The asymptotic behavior of the new caloric curve in 
Fig.\ \ref{fig:twolev} is not an issue, since the integration of Eq.\ 
(\ref{eq:perp}) starts at the saddlepoint, and no local maximum of the 
probability distribution exists. In the lower right-hand panel of Fig.\ 
\ref{fig:twolev}, even the saddlepoint in the probability distribution is not 
well defined, since it depends strongly on the details of the tails of the two 
levels under consideration. Thus, it is harder to uniquely define the new 
caloric curve. If one simply cuts off the tails at some energy, (between 1.8 
and 2.2 in the example), the saddlepoint is completely undefined and the new 
caloric curve has to jump from the lower cut-off energy to the higher cut-off 
energy at a somewhat arbitrary temperature. Although, at a first glance, such a
behavior might imply a first-order phase transition, this is not necessarily 
the case, since additional requirements for the occurrence of a 
phase-transition-like phenomenon in a small system have to be fulfilled 
\cite{BM00,SG02,GC02}. For the sake of the discussion, we have calculated 
several new caloric curves starting all at the cut-off energies but with 
different initial values of the temperature. It can be seen that all curves 
approach the microcanonical caloric curve in the most direct manner and then 
follow it down to zero or up to infinite temperature. Thus, while the 
qualitative features of the new caloric curve are equal for all reasonable 
initial temperatures, the actual transition temperature (where the new caloric 
curve sweeps from $E_1$ to $E_2$) is rather undefined and depends strongly on 
the details of the tails of the two levels under consideration. This fact is 
reflected by the grey-shaded area of Fig.\ \ref{fig:twolev} which covers most 
of the area where reasonably the new caloric curve could be expected.

\section{Conclusions}

It has been shown that traditional caloric curves often fail to capture the
thermodynamics of small systems and describe systems in large regions by 
locally least or even completely impossible values of the energy for a given 
temperature. A different construction for caloric curves in small systems has 
been proposed which mostly omits these drawbacks. Thus, spurious structures 
like negative temperatures and negative heat capacities which are often seen in
the microcanonical caloric curves disappear, but on the other hand, more 
structures in the new caloric curve are evident than in the usually featureless
canonical caloric curves. Although the newly proposed construction of caloric
curves is just that, a construction, it will be interesting to investigate 
whether the more pronounced features of the new caloric curves can be related 
to phase-transition-like phenomena in small systems in a similar way as it has 
been done for the traditional caloric curves in the thermodynamical limit. The 
proposed construction of caloric curves can be applied to study the breaking of
nucleon Cooper pairs where complete level-density information is available. It 
could also be applied for other problems in nuclear physics, e.g., in the 
analysis of multifragmentation data or in the search for the transition to the 
quark-gluon plasma. It is also conceivable that the use of the new caloric 
curves is not limited to atomic nuclei alone, but that it might be applied to 
other small systems in other branches of physics as well.

\acknowledgements

Part of this work was performed under the auspices of the U.S. Department of 
Energy by the University of California, Lawrence Livermore National Laboratory 
under Contract No.\ W-7405-ENG-48. Financial support from the Norwegian 
Research Council (NFR) is gratefully acknowledged. We would like to thank E. 
Tavukcu for carrying out some preliminary calculations for this work. One of us
(A.S.) would like to thank Ben Mottelson, Luciano Moretto and Larry Phair for 
interesting discussions.

\end{multicols}

\clearpage

\end{document}